\documentstyle[preprint,aps,prb]{revtex}
\begin{document}
\draft

\title {\bf Electronic structure calculations  and molecular dynamics
simulations
with linear system-size
scaling
}
\author{ Francesco Mauri and
Giulia Galli
}
\address{Institut Romand de Recherche Num\'{e}rique
en Physique des Mat\'{e}riaux (IRRMA),\\
PHB-Ecublens, 1015 Lausanne, Switzerland.
}
%
%
\maketitle
\begin{abstract}

 We present a method for
total energy minimizations and molecular dynamics simulations
based either on tight-binding or on Kohn-Sham hamiltonians.
The method leads to an algorithm whose
computational cost scales linearly with the system size.
The key features of our approach are
 (i) an orbital formulation with single particle wavefunctions
constrained to be localized in given regions of space,
and  (ii) an energy functional which
does not require either explicit orthogonalization of
the electronic orbitals, or inversion of an overlap matrix.
 The foundations and accuracy of the approach
and  the performances of the algorithm
are discussed, and illustrated
with several numerical examples including Kohn-Sham hamiltonians.
In particular we present calculations with tight-binding hamiltonians
for diamond, graphite, a carbon linear chain and liquid carbon at
low pressure. Even for a complex case such as
liquid carbon -- a disordered metallic system
with differently coordinated atoms --
the agreement between standard diagonalization schemes and our approach
is very good. Our results establish the accuracy and reliability
of the method for a wide class of systems and show that
tight binding molecular dynamics simulations with a few thousand atoms
are feasible on small workstations.
\end{abstract}
\newpage
\narrowtext
\section {Introduction}

 Many studies of materials carried out nowadays in condensed matter physics
are based on total energy calculations and molecular dynamics
simulations with forces derived either from first principles (FP)
or tight binding (TB) hamiltonians \cite{rassegna92}.
These computations rely on a single particle
orbital formulation of the electronic problem.
Within such a framework, the calculation
of the total energy amounts to the solution of a set of eigenvalue
equations (e.g. the Kohn-Sham equations,
in Density Functional Theory), which is obtained
by diagonalizing the
hamiltonian matrix ($H$). $H$ is usually set up according to a chosen basis
set for the electronic orbitals.
Both direct and iterative diagonalizations
imply an overall scaling of the computational effort
which grows as the third power of the number of electronic states,
and thus as the cube of the number of atoms in the systems.
This unfavorable scaling is a major limitation to the use of
TB and FP hamiltonians for systems containing more than a few hundred and
a few thousand electrons, respectively.

 Iterative diagonalizations have been utilized in the study of a variety
of systems in recent years;
indeed when the number $M$ of basis functions is much larger than the number
$N$ of electronic states these schemes are much more efficient than
direct diagonalizations. There are two types of iterative approaches:
constrained minimization (CM) methods  \cite{rassegna92}
in which the single particle wavefunctions
are required to be orthonormal and unconstrained (UM)
methods \cite{GP92,Arias}, in which
the orbitals are allowed to overlap.
In computations with plane wave (PW) basis sets and
pseudopotentials -which are the ones most widely used
in, e.g., first principles molecular dynamics simulations \cite{CP85}-
the evaluation of $\{H\phi\}$, i.e. of $H\phi$ for the $N$
electronic states, requires $O(NM)$ operations
($M$ is proportional to $N$). This is so
if advantage is taken of fast Fourier transform techniques and of the
localized nature of non local pseudopotentials.
The application of orthogonality constraints implies instead $O(N^2M)$
operations. When UM are used, the calculation of the overlap
matrix (${\bf S}$) and of its inverse are of $O(N^2M)$
and $O(N^3)$, respectively.

 Recently several groups have proposed methods to overcome
the problem of the so called $N^3$ scaling, and algorithms
with linear system-size scaling
\cite{Y91,BG91,GP92,WT92,MGC93,LNV93,D93,DS93,K93,ODGM93,A93}.
These approaches are usually referred to as $O(N)$ methods.
Some of them are based on an orbital
formulation of the electronic problem \cite{GP92,WT92,MGC93,K93,ODGM93},
whereas others rely on  formulations without
single particle wavefunctions, but based
on the direct calculation  either of the one electron
Green function \cite{BG91,A93} or of the density matrix
\cite{LNV93,D93}.

 A key idea \cite{GP92} of $O(N)$ orbital schemes is to use
wavefunctions forced to be localized in given regions of space.
These regions are to be chosen appropriately, i.e. large
enough so that the effect of localization constraints be made negligible
on the computed properties.
The solution of the eigenvalue problem by searching directly the
eigenstates is therefore abandoned in favor of a search for a linear
combination of eigenstates which is localized in real space.
In this way the total number of expansion coefficients used to
represent the localized electronic orbitals depends linearly on the
size of the system and the number of operations needed for
the evaluation of $\{H\phi\}$
can be reduced\cite{GP92} to $O(N)$.
 The idea of working with localized wavefunctions is directly related
to that of taking advantage  of the local nature of the density matrix ($\rho$)
in real space \cite{LNV93,D93},
by considering the elements $\rho_{ij}$ to be zero
for distances larger than an appropriate cutoff (localization) radius.

In order to reduce to $O(N)$ operations not only the calculation
of $\{H\phi\}$ but also iterative
orthogonalization
procedures or the ${\bf S}$ inversion, one should in principle resort to
assumptions on the form of the overlap matrix.
If the off diagonal elements of ${\bf S}$
can be made appropriately small, with respect to its
diagonal elements, then the matrix can be inverted
with an iterative procedure whose number of iterations
does not increase with the size of the system, and which therefore implies
a number of operations scaling linearly with system size.
However the problem of imposing {\it explicit} orthogonalization constraints
or of inverting ${\bf S}$ can be solved without
any assumption or approximation.
One can define a functional with {\it implicit} orthogonalization constraints,
containing only the ${\bf S}$ matrix but not its inverse,
in such a way that it has exactly the same minimum as
the Kohn-Sham density functional.
One can therefore use
a functional which is "easier"  both to evaluate and to minimize
than those used in standard CM and UM, which nevertheless
has the "correct" ground state energy and charge density.
This is another key idea
of the $O(N)$ orbital scheme which was introduced in Ref.~\cite{MGC93}.

 Such an approach
will be presented in detail in section II of this paper.
In section III we discuss numerical results obtained for first principles
calculations within density functional theory, in the local density
approximation.
In section IV we demonstrate that an algorithm
with linear system-size scaling can be obtained when the functional
with implicit orthogonalization constraints
is minimized with respect to localized orbitals.
Section V and VI contain our results for the minimization of
tight binding hamiltonians
and for molecular dynamics simulations, respectively.
Summary and conclusions are given in section VI.

\section{ An energy functional with implicit orthogonalization
             constraints }

\subsection{ Definition  and characterization of the energy functional }

 The key points of the unconstrained minimization method
introduced in Ref.~ \cite{MGC93} are:
(i) the replacement of the inverse of the overlap matrix, entering
the energy functional used
in standard UM methods,
with its series expansion in
$({\bf I-S})$ up to an odd order $\cal N$, where
${\bf I}$ is the identity matrix;
(ii) the implicit inclusion of orthonormality constraints
in the energy functional, at variance with standard CM methods, where
orthonormality constraints are treated explicitly, i.e.
as Lagrange multipliers.
After defining the novel energy functional which satisfies
properties (i) and (ii),
we will prove that: (1) this energy functional has the Kohn-Sham
ground state energy ($E_0$) as its
absolute minimum and (2) its minimization yields orthonormal orbitals.

 We consider an energy functional of $N/2$ overlapping orbitals
$\{\phi\}$ expanded in a finite basis set, and of the ($N/2\times N/2$)
matrix ${\bf A}$:
\begin{equation}
E[{\bf A}, \{\phi\}] = 2 ( \sum_{ij}^{N/2} A_{ij}
<\phi_i|-{1\over 2}\nabla^2|\phi_j> +
F[\tilde {\rho}] ) + \eta ( N - \int d{\bf r} \tilde {\rho}({\bf r}) )
\label{energy}
\end {equation}
where
$\tilde{\rho}({\bf r}) = \tilde{\rho}[{\bf A},\{\phi\}]({\bf r}) =
2\sum_{ij}^{N/2} A_{ij} \phi_j({\bf r}) \phi_i({\bf r})$,
$F[\tilde {\rho}]$ is the sum of the Hartree, exchange-correlation
and external potential energy
functionals and $\eta$ a constant
to be specified.
The factor $2$ accounts for the electronic
occupation numbers, which are assumed to be all equal.
For simplicity we consider real orbitals.
According to the choice of the matrix ${\bf A}$, one can obtain either
the functional used in standard UM methods or the energy
functional which we introduced in Ref.~\cite {MGC93}.
 If $ {\bf A} =  {\bf S}^{-1}$,
where $S_{ij}=<\phi_i|\phi_j>$, then
$\tilde{\rho}[{\bf S}^{-1}]$ is the single particle charge density
$\rho ({\bf r})$ and
the term multiplying $\eta$ is zero; in this case
the functional of Eq.~\ref{energy} is the total energy of
interacting electrons in an external field according to
DFT, written for overlapping orbitals \cite{GP92,Arias}.
In particular, if
the wavefunctions are orthonormal then
$ {\bf A} =  {\bf S}^{-1} = {\bf I}$, and
Eq.~\ref{energy} gives the total energy functional of DFT
 used in CM and
ab-initio molecular dynamics (MD) simulations \cite{rassegna92}.
We indicate with $\{ \psi \}$ and $\{ \phi \}$ sets of orthonormal and
overlapping orbitals respectively, and with
$E^{\perp}[\{\psi\}]$ the energy functional of CM
procedures.
The sets $\{ \psi \}$ and $\{ \phi \}$
are related by the
L\"owdin transformation \cite{L50} $\psi_i = \sum_{j} S_{ij}^{-1/2}\phi_{j}$
and then
$E^{\perp}[{\bf S}^{-1/2}\phi] = E[{\bf S}^{-1}, \{ \phi \} ]$.
Therefore
\begin{equation}
\min_{\{\psi\}} E^{\perp}[\{\psi\}] =
\min_{\{\phi\}} E[{\bf S}^{-1},\{ \phi \}]=E_0.
\label {f1}
\end{equation}

The energy functional  of $\{ \phi \}$,
$E[{\bf Q}[\{\phi\}], \{\phi\}]$, which we introduced in Ref.~\cite{MGC93}
is obtained
by taking ${\bf A}$ = ${\bf Q}$ where
\begin{equation}
{\bf Q} = \sum_{n=0}^{\cal N} ({\bf I-S})^n
\label{O}
\end{equation}
and ${\cal N}$ is odd. ${\bf Q}$ is the
the truncated series expansion of ${\bf S}^{-1}$.
We note that similarly to
$E^{\perp}$ and $E[{\bf S}^{-1}]$,
$E[{\bf Q}]$ is invariant under unitary transformations
in the subspace of occupied states, i.e. under the transformation
$\phi^{\prime}_i = \sum_j^{N/2} U_{ij} \phi_j$,
where {\bf U} is a ($N/2\times N/2$) unitary matrix.

 We now prove that the absolute minimum of $E[{\bf Q},\{\phi\}]$ is $E_0$.
If the orbitals are orthonormal, i.e. $S_{ij} = \delta_{ij}$, then
$Q_{ij} = \delta_{ij}$ and $E[{\bf Q}, \{ \psi \} ]$
coincides with $E^{\perp}[\{\psi\}]$:
$
\min_{\{\psi\}} E^{\perp}[\{\psi\}] =
\min_{\{\psi\}} E[{\bf Q}, \{\psi\}]
$.
Furthermore, since $\{\psi\}$ is a subset of $\{\phi\}$,
$
\min_{\{\psi\}} E[{\bf Q}, \{\psi\}] \geq
\min_{\{\phi\}} E[{\bf Q}, \{\phi\}]
$.
As a consequence
\begin{equation}
\min_{\{\psi\}} E^{\perp}[\{\psi\}] \geq
\min_{\{\phi\}} E[{\bf Q}, \{\phi\}].
\label {f2}
\end{equation}
This shows that $E_0$,  is an upper bound
to $\min_{\{\phi\}} E[{\bf Q}, \{\phi\}]$.

 We consider the
difference between the functionals $E[{\bf Q}, \{\phi\}]$
and $E[{\bf S}^{-1}, \{\phi\}]$ i.e.
\begin{equation}
\Delta E = E[ {\bf Q},\{ \phi \} ] - E[ {\bf S}^{-1},\{ \phi \} ] =
\int_0^1 { {\partial E[ {\bf A}(\lambda),\{ \phi \} ]}
\over {\partial \lambda} } d\lambda
\label {variational}
\end{equation}
where ${\bf A}(\lambda) = \lambda ({\bf Q} - {\bf S}^{-1}) + {\bf S}^{-1}$
Using Eq.~(\ref{energy}), Eq.~(\ref{variational}) becomes:
\begin{equation}
\Delta E = 2
\sum_{ij}^{N/2} < \phi_j |
\overline{H}_{KS}
-\eta | \phi_i > ( Q_{ij} - S^{-1}_{ij})
\label {variational3}
\end{equation}
where
$\overline{H}_{KS} =
-{1\over 2}
\nabla^2 + \overline{V}_{KS}$, with
$\overline{V}_{KS}  = \int_0^1 d\lambda
V_{KS} [\tilde{\rho}(\lambda)]$ and $V_{KS} [\tilde{\rho}] =
{ {\delta F}\over {\delta \tilde{\rho} } }$ .
$\overline{H}_{KS}$ is a Kohn-Sham hamiltonian, where
the self-consistent potential is averaged over
the integration path ($\lambda$) of Eq.~(\ref{variational}).
Given a finite basis set for the orbitals $\{\phi\}$, one can choose
$\eta$ large enough so that
the operator ($\overline{H}_{KS} -\eta $)
is negative definite;  then also the ($N/2\times N/2$) matrix
$ <\phi_j | \overline{H}_{KS} -\eta | \phi_i >$ is
negative definite.
 Using the expression of the sum of a geometric series for ${\bf Q}$, we can
express the difference between ${\bf Q}$  and ${\bf S}^{-1}$ as
$({\bf Q} - {\bf S}^{-1}) = -{\bf S}^{-1} ({\bf I-S} )^{{\cal N}+1} =
-({\bf I-S} )^{{\cal N}+1} {\bf S}^{-1}$.
If ${\cal N}$ is odd, the difference between
${\bf Q}$ and ${\bf S}^{-1}$ results to be  a non positive definite
matrix since ${\bf S}$, ${\bf S}^{-1}$ and $({\bf I-S} )^{{\cal N}+1}$ are
commuting non negative definite matrices.
Therefore
if $\eta$ and ${\cal N}$ fulfill the above requirements,
$\Delta E$ is non negative since it is equal to the trace
of the product of a negative and of a non positive definite matrix.
As a consequence,
for each set of $\{ \phi \}$
\begin{equation}
E[ {\bf Q},\{ \phi \} ] \geq E[ {\bf S}^{-1},\{ \phi \} ].
\label{f3}
\end{equation}
The equality holds only if $ ({\bf Q } - {\bf S})^{-1}$
is equal to zero and therefore only if ${\bf S} = {\bf I}$.
Eq.(\ref{f3})
shows that $E_0$ is a lower bound
to $\min_{\{\phi\}} E[{\bf Q}, \{\phi\}]$.
{}From Eqs.~\ref{f1}, \ref{f2} and \ref{f3} we have
\begin{equation}
\min_{\{\psi\}} E^{\perp}[\{\psi\}] = \min_{\{\phi\}} E[{\bf Q}, \{\phi\}] =
\min_{\{\phi\}} E[{\bf S}^{-1}, \{\phi\}] = E_0.
\label{f4}
\end{equation}
This proves that {\it the energy functional $E[{\bf Q}]$ has the Kohn-Sham
ground state energy ($E_0$) as its absolute minimum}, if
$\eta$ and ${\cal N}$ fulfill the requirements discussed above.

 We showed that $E[{\bf Q}]$ and $E[{\bf S}^{-1}]$ are equal only if
the orbitals are orthonormal [Eq.~(\ref{f3})] and that at the minimum
the two functionals are equal [Eq.~(\ref{f4})]. It then follows that
{\it the minimization of $E[{\bf Q}]$ yields orthonormal orbitals}.

The choice of $\eta$ which makes
($ \overline{H}_{KS} -\eta $)
negative definite deserves some comments.
If the hamiltonian of the system does not depend
on $\rho$,
a  $\eta$  larger than the hamiltonian maximum eigenvalue
$\epsilon_{max}$
insures that $\Delta E \geq 0$.
Within LDA, one can prove that
$H_{KS} [\tilde {\rho}[{\bf Q} ]] \leq  H_H [\rho]$, where
$H_H [\rho] = [ -{1\over 2} \nabla^2 + V_{H} [\rho] + V_{ext} ]$,
$V_{H}$ and $V_{ext}$ are the Hartree and external potential,
respectively, and $\rho = \tilde \rho [ {\bf S}^{-1} ]$.
This follows from the property
$\tilde{\rho}[ {\bf Q} ]({\bf r}) \leq
\rho ({\bf r})$,
valid for
each point ${\bf r}$, and from the explicit LDA expression of
the exchange and correlation energy as a function of $\rho({\bf r})$.
Within, e.g., a plane wave
implementation with a finite cutoff, $H_H$ has an upper bound.
This insures the existence of $\eta$ such that
$\Delta E \geq 0$.
 However in practical implementations one can choose $\eta$ smaller
than the upper bound of  $H_H$; indeed
for practical purposes it is not necessary to
require $E_0$ to be the absolute minimum of
$E[{\bf Q}]$, but it is sufficient to require it
to be a local minimum of $E[{\bf Q}]$;
the constant $\eta$ which fulfills this
weaker condition is in general much smaller that the upper
bound of $H_H$, as
we will discuss in the next section.

\subsection { Iterative minimization of the energy functional}

In this section we discuss
the choices of $\eta$ appropriate in
practical applications and the convergence rate of iterative
minimizations of $E[{\bf Q}]$ with ${\cal N} = 1$,
compared to that of $E^{\perp}$. For non self consistent hamiltonians,
we will show that if $\eta$ is larger than the Fermi energy, then
$E_0$ is a local minimum of $E[{\bf Q}]$;
furthermore if a value of $\eta$ is chosen, which is close
to the Fermi energy,
the minimum of
$E[{\bf Q}]$ and that of $E^{\perp}$ can be obtained with
the same computational efficiency.

 The asymptotic convergence rate of iterative minimizations of
 a functional $E[\{\phi \}]$
can be estimated by expanding it around its minimum $E_0$,
up to second order in the variation of the wavefunctions $\{\phi\}$.
As discussed, e.g., in Ref.~\cite{TMC93},
in the  minimization asymptotic regime
the number of integration steps to reach convergence
is directly related to the
ratio between
the maximum and the minimum eigenvalues
of the quadratic form
which results from the second order expansion of $(E - E_0)$.

We consider  a non self-consistent
hamiltonian ($H$) and we relate
its eigenvalues ($\{\epsilon\}$) to those
of the quadratic expansion of $ (E[{\bf Q}] - E_0)$.
Since $E[{\bf Q}]$ is invariant under unitary
transformations in the subspace of occupied states,
a generic variation of the wavefunction with
respect to the ground state can be written as :
\begin{equation}
| \phi_i > =  | \chi^0_i > + | \Delta_i >,
\label{displac}
\end{equation}
with
\begin{equation}
| \Delta_i > =  \sum_{l\in [\cal{TOT}]}  c^i_l | \chi^0_l >.
\end{equation}
Here
$| \chi^0_l >$ are the eigenstates of $H$,
and the indeces $i$ and $l$ belong to the set
of occupied states and to the set of occupied plus empty
states, respectively.
We denote with [$\cal {OCC} $] and [$\cal {EMP}$] the sets of occupied
and empty states, and with [$\cal {TOT}$] the union of the two
sets.
$c^i_l$ are expansion coefficients of
$ |\Delta_i>$ over the eigenstates of $H$.
If Eq.~\ref{displac} is substituted into the expression of
$\Delta E = E[{\bf Q}] - E_0$, the first order term vanishes,
showing that for each value of $\eta$ the orbitals $ | \chi^0_i > $ make
$E[{\bf Q}]$ stationary. The stationary point is
in particular an absolute minimum if $\eta \geq
\epsilon_{max}$, as shown in the previous section.
One is then left with a second
order term, which can be recast as follows:
\begin{equation}
 \Delta E =   \sum_{m\in [{\cal EMP}]}  \sum_{i\in [{\cal OCC}]}
2 [\epsilon_m - \epsilon_i] (c^i_m)^2 +
\sum_{ij\in [{\cal OCC}]} 8 [ \eta - { {(\epsilon_i+\epsilon_j)}\over {2} } ]
[{1\over \sqrt{2}} (c^i_j + c^j_i)]^2
\label{eigenmodes}
\end{equation}
{}From Eq.~\ref{eigenmodes} it is seen that the quadratic form $\Delta E$
has two sets of normal modes. The first set has eigenvalues
$k_{(mi)} = 2 [\epsilon_m - \epsilon_i]$,
which are always positive and independent of $\eta$; they correspond
to the coordinates $c^i_m$. These modes
are associated
with an increase of the total energy when the orbitals acquire non zero
components on empty eigenstates of $H$.
They are the same as the normal modes of $(E^{\perp} - E_0)$,
calculated in Ref.~\cite{PSB91}.
The second set of normal modes of $\Delta E$ has eigenvalues
$\overline{k}_{(ij)} = 8 [ \eta - { {(\epsilon_i+\epsilon_j)}\over {2} } ]$;
they correspond to the coordinates $[{1\over \sqrt{2}} (c^i_j + c^j_i)]$.
These modes are associated with a change of $E[{\bf Q}]$ due to the overlap
of the electronic wavefunctions; they are indeed associated with
the orthogonality constraints implicitly included in the
definition of $E[{\bf Q}]$ and they are not present
in $(E^{\perp} - E_0)$.

 For $\eta$ larger than the
highest occupied eigenvalue of $H$, $\epsilon_{N/2}$ (i.e. the Fermi energy),
the $\overline{k}_{(ij)}$ are positive and thus
$E_0$ is a local minimum of $E[{\bf Q}]$.
$\eta \geq \epsilon_{N/2}$ is a weaker condition than the one required
to prove Eq.~(\ref{f4});
it is however a sufficient condition to insure that the
minimization of $E[{\bf Q}]$ leads to the
correct ground state energy, provided
a reasonable starting point for the minimization is chosen.
This will be shown also with numerical examples in the next section.

The minimizations of
$E[{\bf Q}]$ and $E^{\perp}$ can be obtained
with the same efficiency
provided the weaker condition on $\eta$
is adopted.
For example one can choose $\eta \simeq \epsilon_{N/2 + 1}$.
Under such a condition
the ratio between the maximum and the minimum eigenvalues
of the expansion of $(E[{\bf Q}] - E_0)$ and of $(E^{\perp} - E_0)$
is the same in most systems.
Indeed the eigenvalues $\overline k$
lie in the interval defined by the eigenvalues $k$, if
the spread in energy of the
excited states of $H$ is four times smaller than the valence band width.
This condition is satisfied in most systems of interest.
This means that
in practice iterative minimizations of $E[{\bf Q}]$ and MD
simulations with $E[{\bf Q}]$ can be performed with the same
efficiency as the corresponding calculations with $E^{\perp}$.
However, if $\eta$ is chosen so that
$E_0$ is
an absolute minimum of $E[{\bf Q}]$, the time step used in MD simulations,
which is proportional\cite{PSB91,TMC93} to the square root of the maximum
eigenvalue of $\Delta E$ (Eq.~\ref{eigenmodes}),
is reduced by a factor of two with respect
to that used in standard calculations.

The functional introduced in II.A
has clear advantages over standard
energy functionals when
conjugate and preconditioned conjugate gradient
minimization procedures are used:
the complication of imposing orthonormality constraints is avoided,
and contrary to ordinary unconstrained methods an automatic
control of the ${\bf S}$ matrix is provided,
since at the
minimum ${\bf S}  = {\bf I} $.
Furthermore, when preconditioning of the high frequency
components of the single particle wavefunctions is introduced,
e.g. in Car-Parrinello molecular
dynamics simulations, the integration of the electronic equation
of motion does not imply
any extra work, at variance with integration schemes
with explicit orthogonalization constraints \cite{TMC93}.

\subsection {Relationship with other functionals}

 The total energy minimization scheme
which we introduced in Ref.~\cite{MGC93} is related to
other approaches recently proposed in the literature for
electronic structure calculations with linear system-size scaling.
In particular Ordej\'on et al.~\cite{ODGM93} derived the same
functional as that of Eqs.~1 and 3 for ${\cal N}$ = 1 for
non self-consistent hamiltonians.
Their derivation is based on a Lagrangian formulation
with explicit orthogonalization constraints, where the
Lagrange multipliers ($\lambda_{ij}$)
are approximated by
an expression which is exact only at the minimum, i.e.
$\lambda_{ij} = <\phi_i | H | \phi_j>$.
 The approach presented by Ordej\'on et al.~\cite{ODGM93}
is similar to that of Wang and Teter \cite{WT92},
although in Ref.~\cite{WT92}
constraints are introduced by means of a penalty function.
However the minimum of the Wang and Teter functional is $E_0$ only if
the weight of the penalty function goes to infinity, at variance with
our and Ordej\'on et al.'s functionals which at the minimum
is always equal to $E_0$.

 Instead of using an orbital formulation, Li, Numes and Vanderbilt
 \cite{LNV93} and Daw \cite{D93}
proposed  a functional for total energy minimizations
within a density matrix formulation.
In this case one minimizes the energy functional with respect
to the density matrix,
which must fulfill the idempotency condition.
This condition is enforced by minimizing the total energy
with respect to a {\it purified} version of the density matrix
\cite{LNV93,mcW72}
($\tilde{\rho}(r,r')$), constructed from a trial density matrix
$\rho (r,r')$ in such a way that
its eigenvalues lie  on the interval [0,1].
The energy functional $E[{\bf Q}]$ (Eq.~1,3) for non self-consistent
hamiltonians
can be re-derived within the formulation of Ref.~\cite{LNV93} if $\rho (r,r')$
is expressed in terms of the occupied single particle wavefunctions, i.e.
$\rho (r,r') =  \sum_{i \in [\cal{OCC}]} \phi_i(r) \phi_i(r')$,
and a purification transformation is chosen such that
$\tilde{\rho} =  I - (I - \rho)^{ {\cal N} + 1}$.
This transformation forces the eigenvalues of  $\tilde{\rho}$ to be less
than 1 only if ${\cal N}$ is odd;
one does not need to force the eigenvalues to be
positive, as done in Ref.~\cite{LNV93}, since by construction
$\rho (r,r') =  \sum_{i \in [\cal{OCC}]} \phi_i(r) \phi_i(r')$
has a number of non zero eigenvalues equal to the number of occupied states.

\section{ Numerical results of first principles calculations}

 The validity of the minimization scheme presented in Section II
was tested numerically for KS hamiltonians within LDA, by computing
the ground state energy of
Si in the diamond structure.
We used an expansion coefficient ${\cal N} = 1$ to define the
${\bf Q}$ matrix entering the energy functional (see Eq.~3).
We chose $\eta$ smaller that the maximum eigenvalue of
$H_{KS}$; this choice insures the iterative minimization
to properly converge to the ground state energy E$_0$, unless
a pathological starting point for the electronic orbitals is chosen.

$E[{\bf Q}]$ was minimized by
steepest descent;
the derivative
of the functional with respect to the single particle orbitals is
given by:
\begin{equation}
{ {\partial E[{\bf Q}]} \over {\partial \phi_i} }=
4 \sum_j^{N/2} [ (H_{KS} -\eta )| \phi_j > (2\delta_{ji} - S_{ji}) -
                        | \phi_j > < \phi_j| (H_{KS} - \eta) | \phi_i > ]
\end{equation}
The orbitals were expanded in PW with a kinetic energy cutoff ($E_{cut}$)
of 12 Ry and the interaction between ionic cores and valence electrons
was described by  a norm conserving pseudopotential \cite{bhs82} expressed in a
separable form \cite{kb82}.
The calculation was started from orbitals set up from random
numbers, with $\eta$ set at 3.0 Ry above the top of the valence band.
In Fig.~1 we report $E^{\perp}$ and $E[{\bf Q}]$ as a function of the
number of iterations;  it is seen
that the minimizations of the two functionals require the same
number of iterations and leads to the same energy.
Fig.~2 shows the integral of the charge density during the minimization
procedure.
For $\cal N$ = 1, $\Delta N = N - \int d{\bf r} \tilde {\rho}({\bf r})
= N - Tr({\bf Q} {\bf S})$
is given by
\begin{equation}
\Delta N = Tr(({\bf I} - {\bf S})^2).
\end{equation}
This is a positive quantity which goes to zero as the orbitals
become orthonormal. In our calculation the difference $\Delta N$
between the total number of electrons and the integrated charge
reaches a value very close to zero ($\simeq 10^{-6}$)
 after 10 iterations, showing that the single
particle wavefunctions are orthonormal already well before
reaching the minimum.
\section{ Localized orbitals and
 an algorithm with linear system-size
scaling}

We now turn to the discussion of the approach introduced in section II
within a localized orbital (LO) formulation \cite{GP92}.
Within such a formulation, each single particle wavefunction
is constrained to be localized
in an appropriate region of space, which we call localization region
(LR): the electronic orbitals are free to vary inside and
are zero outside the LR.
Different single particle orbitals can be associated to the same LR,
e.g. two doubly occupied orbitals per LR for  C and Si,
which have four valence electrons.
The extension of a LR is determined by the
bonding properties of the atomic species composing the
system, and it is the same unrespective of the size of the
system which is simulated.
The choice of the centers of the LRs is arbitrary. In all of our calculations
(see next section) we centered the LRs on atomic sites; this choice
is physically unbiased, i.e. it can be adopted for a generic system
whose bonding properties are totally unknown, and allows for a solution
which satisfies charge neutrality conditions.
If one wants to take advantage of known properties of the system,
LRs can for example be centered on atomic bonds or on positions
compatible with the symmetry of
the Wannier functions, if these can be defined.
This is however difficult to do, e.g. at each step of a MD simulation,
where the evolution of the bonding properties as a function
of time is not known.
One could also treat the centers of the LRs as variational parameters
and optimize their locations during the calculation.

We now consider the minimization of $E[{\bf  Q}]$ with respect
to LO ($\{ \phi^L \}$).
When the orbitals are localized,
$S_{ij}$, and $<\phi^L_i | H_{KS} |\phi^L_j>$ are
sparse matrices which have non zero elements
only if $i$ and $j$ belong to overlapping LRs.
The evaluation of $E[{\bf Q}]$ (Eqs.~1,3) as well as of
${ {\partial E[{\bf Q}]} \over {\partial \phi^L_i} }$
(Eq.~12) implies only the calculation of
matrix products containing $S_{ij}$ and
$<\phi^L_i | H_{KS} |\phi^L_j>$. No
orthogonalization or ${\bf S}$ inversion is needed.
Thus at each step, the minimization of $E[{\bf  Q}]$ can be performed
with a number of operations which is proportional to the system size.

When localization constraints are imposed,
the variational freedom of the minimization procedure
is reduced.
The energy obtained by minimizing  a functional
with respect to LOs is then larger than the absolute minimum  ($E_0$)
obtained with no constraints on the single particle wavefunctions.
In particular, the minimum of $E[{\bf Q}]$ with respect to
LO $\{ \phi^L \}$
does not coincide with  that of
$E[{\bf S}^{-1}]$,
and the LO which minimize $E[{\bf Q}]$
are in general not orthonormal.
This is easily seen as follows.
Whereas Eq.~\ref{f2} and \ref{f3} hold also for LO,
Eq.~\ref{f1} is no longer valid when localization constraints are imposed.
Indeed the transformation from $\{ \psi \}$ to $\{ \phi \}$ with
${\bf S}^{-1/2}$ does not preserve the size of the LR, i.e.
it does not map functions localized in a given region onto functions
localized in the {\it same} region of space.
Therefore Eq.~\ref{f4} does not hold
but is replaced by
\begin{equation}
\min_{\{ \psi^L \}}E^{\perp} \geq
\min_{\{ \phi^L \}}E[{\bf Q}] \geq
\min_{\{ \phi^L \}}E[{\bf S^{-1}}]
\geq E_0,
\label{loc1}
\end{equation}
where the LR for the $\{ \psi^L \}$ and $\{ \phi^L \}$ are the same.
Since in Eq.~(\ref{loc1})
the equality is in general not satisfied,
at the minimum ${\bf S}$ is different from ${\bf I}$,
contrary to the case of extended orbitals.

 The variational quality of the results obtained by minimizing
$E[{\bf Q}]$,
i.e. the difference $[\min_{\{ \phi^L \}}E[{\bf Q}] - E_0]$,
 depends upon  (i) the order ${\cal N}$ chosen for the
definition of the ${\bf Q}$ matrix and (ii) the size of the
LR.
For ${\bf S}\leq 2 {\bf I}$, it is easy to see that
$E[{\bf Q}({\cal N} -2)] \geq E[{\bf Q}({\cal N})]$. Therefore
by increasing
${\cal N}$ in the definition of ${\bf Q}$,
one obtains an improvement of the total energy. This
leads as well to an increase
of the number of operations needed
in the computation of ${\bf Q}$ (see Eq.~3).
Most importantly,
in order to improve the quality of the results
one can choose to increase the size of the localization
region.
We note that the number of non zero elements of ${\bf S}$ is proportional to
$n_{LR}N$, where $n_{LR}$ is the average number of regions overlapping
with a given one. Instead the number of degrees of freedom needed
to define the $N/2$ single particle orbitals is proportional to
$mN$, where $m$ is the number of points belonging to a LR, e.g. the
number of points where the wavefunction is non zero.
The ratio $n_{LR}/m$ strongly depends on the basis set chosen to set up
the hamiltonian. The optimal choice of ${\cal N}$
and of size of the LRs, e.g. of the parameters determining the
efficiency and accuracy of the
computation, crucially depends upon the chosen basis set.

In calculations where $m \gg n_{LR}$, the computer time for the ${\bf S}$
inversion amounts to a small fraction of the total time also
for relatively large systems (e.g. systems with up to
 a few thousand electrons in LDA calculations with PW basis).
On the contrary for computations
with small basis sets, such as those with TB hamiltonians,
the computer time for the ${\bf S}$ inversion constitutes a
considerable part of the total time already
for  small systems (i.e. containing a few tens of atoms).

\section{ Minimization of TB hamiltonians}

 The LO formulation was tested numerically
using TB hamiltonians \cite{gsp89,xwch92} with the convention
$\varepsilon_s+\varepsilon_p=0$.
We performed calculations for Si and C in different aggregation states.
In calculations
for crystalline structures, we considered non zero hopping terms
only between first neighbors.
We chose a number of LRs equal to the
number of atoms and we centered each LR at an atomic  site ($I$).
In a TB picture a LR can be identified with the
set of atoms belonging to it. For each site
$I$, we label the set of atoms which belong to a LR with
$LR_{I}$. C and Si atoms have four valence electrons and
there are two doubly occupied states for each atom in the
system.
We then associated two states to each LR:
The two wavefunctions of the LR centered in $I$ have non
zero components on the atoms belonging to the set $LR_{I}$ and
zero components (expansion coefficients)
on the atoms which do not belong to $LR_{I}$.
The expansion coefficients of the single particle orbitals
are treated as variational parameters in our calculations.
The total number of expansion coefficients grows linearly with the size of
the system.

 We tested two different shapes of the LR.
In one case an atom is defined as belonging to $LR_{I}$ if its distance
to the site $I$ is less than or equal to a given radius $r_c$ (in other words,
an euclidean metric is used to define the shape of the LR).
In the second case, we took
advantage of the form the TB hamiltonian and we considered
an atom as belonging to $LR_{I}$
if it is connected to the site $I$ by a number of non zero
hopping terms less than or equal to a given number of shells $N_{h}$.

In all calculations $E[{\bf Q}]$ was minimized
with respect to $\phi_L$
by a conjugate
gradient (CG) procedure. The gradients
$ {{\partial E{[\bf Q}]} \over {\partial
\phi^L_i} }$ are simply obtained by projecting Eq.~(12) onto the LR
where $\phi^L_i$ is defined.
For non self-consistent hamiltonians, the line
minimization required in a CG procedure reduces to the minimization of
a quartic polynomial in the variation of the wavefunction,
along the conjugate direction. In our calculation the line minimization
is performed exactly by evaluating the coefficients of the quartic polynomial.

We found that when localization constraints are imposed, $E[{\bf Q}]$
can have local minima
and metastable states, where the system may be trapped for a
long time during the minimization procedure, before reaching a minimum.
This problem can be overcome if an appropriate choice of
the initial guess for the iterative diagonalization is made.
In all of our calculations we used starting wavefunctions
with non zero components only on
the site $I$ where they were centered; furthermore orbital components
were the same for each $I$.
This choice allowed
to avoid local minima and metastable state traps
for a wide class of ionic configurations.
The problem of being trapped in metastable states or local
minima involves only electronic minimizations; it does not concern
MD simulations, where the ground state orbitals of a given step
can be used as guess wavefunctions for the following step.

  Fig.~3 shows the percentage error on the cohesive energy
$E_c$ of Si in the diamond structure,  as a function
of the size of the LR, computed with respect to a calculation
perfomed with extended orbitals. All computations were carried out
with 216 atom supercells, simple cubic periodic boundary conditions and
the $\Gamma$ point only for the supercell Brillouin zone (BZ) sampling.
$E_c$ was evaluated with
${\bf Q}[{\cal N}= 1]$
and ${\bf Q}[{\cal N}= 3]$ and with $\eta$ = 3 eV.
The shape of the LR was first chosen using an euclidean metric.
We denote with $N_e$ the number of shells included in a LR,
defined according to such
a metric. It is seen that $E_c$ converges rapidly as a function of
$N_e$, with both ${\cal N}=1$ and $3$. Already with $N_e=2$ (17 atoms
belong to a LR) the results are very good,
i.e. $E_c$ is higher
than the result obtained with extended orbitals by only $2.1\%$
and $0.8\%$ for ${\cal N}=1$ and $3$, respectively.
For ${\cal N}=1$, the error on the total charge
$\Delta N$ (see Eq.~13) which gives the deviation from
orthonormality due to localization constraints is in general
very small; already for $N_e=2$ we find it to be $0.2\%$.
We note that when going from
$N_e=3$ (29 atoms in a LR) to $N_e=4$ (35 atoms in a LR),
we obtain the smallest variation of $E_c$. Indeed
the atoms added to a LR when
including also the fourth neighbor shell are
not connected by hopping terms to those
defining a LR when $N_e=3$. This suggests that a
definition of LR based on hopping terms is more physical than
one based on the euclidean metric.
We repeated the calculations with ${\cal N}=1$ by choosing
the LRs according to the hopping parameters and by setting
the number of hopping shells
$N_h$ at 3. (For the diamond lattice, the definition
of LRs using the two metrics are different for $N_h$ and $N_e$ larger than
two).
The choice $N_h=3$ amounts to considering 41 atoms in
a LR. The percentage error (0.7 $\%$) on  $E_c$
is very close to that obtained with $N_e=5$
(0.6 $\%$), although the
number of atoms in a given LR is bigger (47).
 The choice of the shape of the LRs according to the
hopping parameters is superior to that of
the euclidean metric  and it is especially so when energy
differences between different structures are to be computed.
This is the definition which was adopted in all subsequent calculations
for C.

 Results for carbon in different crystal structures are presented
in Tables I and II and in Fig.~4.  We chose systems with
different bonding and electronic properties: a sp$^3$
bonded insulator, diamond, a sp$^2$ bonded semi-metal, planar graphite,
and a sp bonded metal, a non dimerized C chain.
Table I shows the binding energy of the three structures
as  a function of the size of the LR.  The calculations were performed with
$E[{\bf Q}({\cal N}= 1)]$. The errors for $N_h=2$ and $N_h=3$ are of the
same order of those found in the case of silicon, and in particular
we find that already for $N_h=2$ the LO formulation
and  a direct diagonalization  scheme are in good accord.
In Fig.~4 we compare the total energy of the three C systems
as a function of the lattice parameter,
as obtained by direct diagonalization of  the hamiltonian and by
minimizing $E[{\bf Q}({\cal N}= 1)]$ with respect to LO, with $N_h = 2$.
The agreement between the two calculations is again very good for the
three systems, in spite of their
different bonding and electronic properties.
The percentage difference between the computed
equilibrium properties (lattice constant, cohesive energy and
bulk modulus) are given in Table II.


\section{ Molecular Dynamics with TB hamiltonians}

By using the functional $E[{\bf Q}]$ and localized orbitals one
can set up a MD scheme in which the computational cost
of each step scales linearly with the system size.
According to the Helmann-Feynman theorem,
one can obtain the forces acting on a given atom $I$
by computing ${\bf F}_I = - \nabla_I E[{\bf Q}; \{\phi_L\}, \{ {\bf R}_I \}]$;
here
(${\bf R}_I$) denotes ionic positions and $\{\phi_L\}$ are
the localized orbitals which minimize $E[{\bf Q}]$.
The general expression of the ionic forces is given by
${\bf F}_I = -  2 \sum_{i,j}^{N/2} Q_{ij} < \phi_i |
{ {\partial V} \over {\partial {\bf R}_I } } | \phi_j >$,
where $V$ indicates the external potential in a LDA calculation and the
hamiltonian in a TB calculation.
In practical computations it is convenient to first calculate the auxiliary
wavefunction
\begin{equation}
| \overline{\phi_i} > =
\sum_{j}^{N/2} Q_{ij} | \phi_j >
\end{equation}
and then to evaluate the expression of ${\bf F}_I$ as follows:
\begin{equation}
 {\bf F}_I = - 2 \sum_{i}^{N/2}
< \phi_i |{ {\partial V} \over {\partial {\bf R}_I } } | \overline{\phi_i} >.
\end{equation}

 The ground state wavefunctions $\{\phi_L\}$ can be obtained either
by evolving the electronic states according to a
Car-Parrinello \cite{CP85} dynamics  (see, e.g. Ref.~\cite{MGC93}), or
by minimizing the energy functional $E[{\bf Q}]$ at each ionic move.
In our simulations,  we determine the
sets $LR_I$ at each ionic step; consequently
the sites belonging to a set vary as a function of time, when, e.g.
the atoms are diffusing or
changing their local coordination.
This implies an abrupt modification of the basis functions
used for the expansion of $\{\phi_L\}$ and therefore a discontinuity of
$\{\phi_L\}$ as a function of the ionic positions.
In correspondence to any change of the sets $LR_I$, $E[{\bf Q}]$ must
be minimized with respect to the electronic degrees of freedom;
we therefore chose to minimize the energy functional
at each ionic step, unrespective of whether the LR
changes at a given step.  The minimizations were performed with
a conjugate gradient procedure
where we used as initial guess for the orbitals the linear
extrapolation of the minimized wavefunctions of the two previous
ionic steps, as
suggested in Ref.~\cite{Arias}.

In order to test the accuracy and efficiency
of the LO scheme for different classes of systems,
we performed MD simulations for a crystalline
insulator, i.e. diamond at low temperature, and for a liquid metal,
i.e. liquid carbon at $T\simeq 5000 K$.
As for the calculations for C presented in the
previous section, we adopted LRs
centered on atoms,  which
include up to second shell of neighbors and whose shape is
determined by the hopping parameters.

 We first discuss the case of crystalline diamond, when
the sets $LR_I$ do not vary in time.
We find that for  diamond our MD scheme allows for a correct
description of the total energy oscillations, around equilibrium,
consistently with what previously obtained \cite{MGC93} for Si.
We performed two simulations, one with a 64 atom and the other with a 1000 atom
supercell. In both cases we started from a ionic configuration
with zero velocities, generated by giving
a random displacement to the atoms
up to .03 \AA\  with respect to their equilibrium positions.
The integration time step ($\Delta t$) used in the simulations was 30 a.u.
and the number of CG iterations per ionic move  was 10.
In Fig.~5 we show the potential energy  ($E$) and the sum of the kinetic
($E_{kin}$) and potential
energy of the system as a function of the simulation time. It is seen
that the same energy  drift $\Delta (E + E_{kin}) / E_{kin}$ (0.1 in 0.5
ps) was found for the two simulations. This shows that the
number of CG iterations to obtain a given accuracy in the energy
conservation does not depend on the size of the
system and that the overall scaling of the computational scheme is
therefore linear.
Finally we evaluated the relative error on the ionic forces
${\bf F}_I$ introduced by localization
contraints as $
 { \Delta  F \over F}
 =  { \overline
{\sum_I | {\bf F}_I^{loc} - {\bf F}_I^{ext} | }
\over
\overline {  \sum_I |{\bf F}_I^{ext}| }}$,
where the overline indicates time averages, and the upperscripts
$loc$ and $ext$ refer to calculations performed with localized and extended
states, respectively. This error was found to be $\simeq 6 \%$ in crystalline
diamond at room temperature.

 We note that if extended states are used,
the number of iterations needed to have the same
conservation of energy  as the one reported in Fig.~5
is smaller than 10.
Nevertheless our MD scheme applied to ordered systems becomes more efficient
than direct diagonalization of the hamiltonian already for small systems,
i.e. for systems containing more than 40 atoms.
This can be seen in Fig.~6 where we compare the efficiency of our approach
to that of direct diagonalization based MD schemes.

 We now analyze a MD simulation run during which the sets $LR_I$
change as a function of time. In Fig.~7 we show the potential energy
for an oscillation of crystalline diamond around equilibrium,
computed with extended (E$^{\rm ext}$, dotted line)
and with localized (E$^{\rm loc}$,solid line)
orbitals as a function of simulation
time ($t$). The two energies have been computed for the same ionic
trajectories,
generated by a simulation with localized orbitals.
The MD run shown in Fig.~7 is the same as the one reported in Fig.~5 but
now the LRs are allowed to vary in time.
At $t$ = {\bf t1}, the evolution of the ionic positions makes
the number of atoms belonging
to given localization regions to increase. At $t$ = {\bf t2},
the ionic configuration is such to restore the localization regions
as they were at $t \leq$ {\bf t1}.
Since at $t$ = {\bf t1}, {\bf t2} an
abrupt modification of the basis functions
used for the expansion of $\{\phi_L\}$ occurs, the potential energy
E$^{\rm loc}$ is discontinous and its derivative with
respect to ionic positions is not well defined. However ionic forces can
still be defined by  neglecting the discontinuity in E and by evaluating
either the left or the right derivatives of the potential energy.
The numerical values of the left and right derivatives
are in fact the same within a very small error.
This error is negligible, being much smaller
than the one introduced by localization contraints.
 This can be seen in Fig.~8 where we compare forces obtained
in calculation with extended  and localized orbitals by plotting
dE$^{\rm ext}$/dt = $\sum_I {\bf F}_I^{\rm ext} \cdot {\bf v}_I$
(dotted line) and
dE$^{\rm loc}$/dt = $\sum_I {\bf F}_I^{\rm loc} \cdot {\bf v}_I$ (solid line).
On the scale of the picture no dicontinuity is
observable in dE$^{\rm loc}$/dt at $t$ = {\bf t1}, {\bf t2}.

 We now turn to the discussion of the simulation of liquid C, during which
many changes of $LR_I$ were observed.
We generated a diffusive state at
$T \simeq 5000$  K starting from a diamond network prepared at a macroscopic
density of 2 grcm$^{-3}$; we then heated the system by means of
a Nose'-Hoover thermostat.
We used a 64 atom cell with simple cubic periodic boundary conditions and
only the $\Gamma$ point to sample the BZ. We used a cutoff radius
of 2.45 \AA\  for the hopping parameters entering the
TB hamiltonian and for the two body repulsive potential \cite{xwch92}
(i.e. the cutoff distances r$_m$ and d$_m$ of Ref.~[22] are set at  2.45 \AA\
).
Equilibration of the system was performed in the canonical ensemble
and temporal averages were taken over 3.8 ps.
The same simulation was repeated twice: once with our MD scheme and
once by using direct diagonalization at each step.
The radial distribution function $g(r)$  and the partial
atomic coordinations obtained in the two cases are
shown in Fig.~9 and Table III, respectively.
The agreement between the two descriptions is excellent, showing that
the LO scheme is accurate
even for a difficult case such as a
disordered system with differently coordinated
atoms and metallic properties.
The self-diffusion coefficients obtained in the two cases
are 0.4 10$^{-4}$ cm$^2$sec $^{-1}$ and
0.6 10$^{-4}$ cm$^2$sec $^{-1}$, respectively.
The difference between the cohesive energies computed within the
extended orbitals and the LO formulation for given ionic
configurations is of the order of 2$\%$, similarly to
what found for crystalline structures.

 In the simulation for the liquid with LO, we used
$\Delta t = 5 $ a.u. and we performed $50$ iterations
per ionic move, in order to minimize $E[{\bf Q}]$.
This number is much larger than that needed for ordered systems,
such as crystalline
diamond. Consequently in the case of liquid C our scheme becomes advantageous
with respect to direct diagonalization when the number of atoms
is larger than $200$.

\section{ Conclusions }

We have presented an  approach  to total energy
minimizations and molecular dynamics simulations whose
computational workload is linear as a function of the system size.
This favourable scaling is obtained by using an energy functional
whose minimization does not imply either explicit orthogonalization of the
electronic orbitals or inversion of an overlap matrix, together with
a localized orbital formulation.
The use of LOs reduces the evaluation of the energy functional
and of its functional derivative to
the calculation of products of sparse matrices.

 The performances and efficiency of the method have been illustrated
with several numerical examples for semiconducting  and metallic systems.
In particular we have presented molecular dynamics simulations
for liquid carbon at 5000 K, showing that even for  the case
of a disordered metallic system the description
provided by the LO formulation is reliable and very accurate.
We have also shown that tight binding molecular dynamics simulations
with 1000 atoms are  easily feasible on small workstations, implying
a one day run to obtain 0.5 ps.
Molecular dynamics simulations for very large C systems are underway.

\section{ Acknowledgement}

 It is a pleasure to thank R.~Car, R.~M.~Martin and
D.~Vanderbilt for useful discussions and
A.~Possoz for her help in the optimization of the computer codes.
We acknowledge support by the Swiss National
Science Foundation under grant No. 21-31144.91.
%
%
%

%
\newpage

\figure { {\bf Fig.~1 }
Total energy  (E) as a function of the number of
iterations for a steepest descent minimization of 64 Si atoms in
the diamond structure, described
within LDA with a PW basis set.
The solid and dotted lines correspond
to the minimization of $E[{\bf Q}]$
and $E^{\perp}$ (see text), respectively.
${\bf Q}$ was defined with ${\cal N} = 1$.
We used kinetic energy cutoffs of $12$  and $36$ Ry for the
wavefunctions and charge density, respectively
and  we set $\eta $  at 3 Ry above the top of the valence
band. Each run was started from the same set of random Fourier coefficients.
}

\figure { {\bf Fig.~2 }
Total electronic charge
as a function of the number of iterations for the energy
minimizations reported in Fig.~1. The total number of electrons in the system
is 256.
}

\figure { {\bf Fig.~3 }
 Percentage error on the cohesive energy
of Si (diamond structure) as a function
of the number of shells ($N_e$) in the localization region, computed with
a TB hamiltonian (see text).
Diamonds and crosses refer to minimizations of
$E[{\bf Q}]$ with ${\bf Q}[{\cal N}= 1]$ and ${\bf Q}[{\cal N}= 3]$,
respectively.
The LRs were defined using an euclidean metric (see text).
The errors  were evaluated with respect to a computation
with extended orbitals. Calculations were performed at the
same fixed volume.

\figure { {\bf Fig.~4 }
Total energy  ($E$) of diamond (dots), bidimensional graphite (open circles)
and a carbon linear chain (squares) as
a function of interatomic distance ($d$), computed with a TB hamiltonian and
supercells containing 216, 128 and 100 atoms, respectively.
The dotted lines were obtained
by minimizing at each volume
 $E[{\bf Q}]$ with ${\bf Q}[{\cal N}= 1]$,
and by using localization regions defined with $N_h=2$.
The solid lines were instead
obtained by diagonalizing the hamiltonian, with
no constraints on the wavefunctions.
}

\figure{ {\bf Fig.~5}
Potential energy (lower part) and  the sum of the potential and kinetic
energy (upper part) as a function of simulation time for
crystalline C in the diamond structure at  70 K.
The dotted and solid lines refer to two calculations performed with a TB
hamiltonian, with  64 atom and  1000 atom supercells, respectively.
In both cases we used ${\bf Q}[{\cal N}=1]$ and $N_h$ = 2; the LRs
were computed for the configuration at 0 K and held fixed during
the whole simulation.
}

\figure{ {\bf Fig.~6 }
CPU time per ionic step (30 a.u.)  as a function of the
number $N$ of atoms in the system,  for a TB-MD simulation of
C diamond at low temperature (see text). Squares and crosses
indicate the CPU time in
a direct diagonalization based scheme
and in our MD approach (with 10 CG iterations per ionic move), respectively.
Calculations were carried out on a Silicon Graphics Iris Indigo 4000.
}

\figure{{\bf Fig.~7}
 Potential energy
for an oscillation of crystalline diamond around equilibrium,
computed with extended (dotted line)
and with localized (solid line) orbitals as a function of simulation
time. The two energy curves have been computed for the same ionic trajectories,
generated by a simulation with localized orbitals.
The LO calculation is the same as the one carried out in Fig.~5, but
here the LRs are allowed to vary during the simulation.
{\bf t1} and {\bf t2} denotes times at which the LRs change.
}

\figure{{\bf Fig.~8}
Time derivatives
dE$^{\rm ext}$/dt = $\sum_I {\bf F}_I^{\rm ext} \cdot {\bf v}_I$
(dotted line) and
dE$^{\rm loc}$/dt = $\sum_I {\bf F}_I^{\rm loc} \cdot {\bf v}_I$ (solid line)
of the potential energy curves reported in Fig.~7 (see text).

\figure{ {\bf Fig.~9 }
Radial distribution function g(r) of liquid C (see text) computed
as average over a TB-MD simulation of 3.8 ps.
The results of the LO formulation with $N_h = 2$ and ${\cal N}=1$ (dotted line)
are compared to those of a direct diagonalization scheme
(solid line). The average number of atoms in a LR is 18.
}
%
%
\begin{table}
\begin{tabular}{|l|c|c|c|c|}
          &         &                      &  &              \\
Crystal structure   & {\bf $ r_0$}
&  {\bf  E$_c$ [$N_h=2$] }  &  {\bf E$_c$ [$N_h=3$] }&
{\bf  E$_c$ [$N_h= \infty $] }     \\
          &         &                      &  &              \\ \hline
          &         &                      &  &              \\
Diamond & 1.54    &7.16 &7.23& 7.26\\
2D-graphite & 1.42&7.09 &7.19&7.28\\
1D-chain & 1.25   & 5.62&5.75&5.93\\
          &                         &      &    &            \\
\end{tabular}
\vskip 0.5truecm
\caption{
Cohesive energy E$_c$ (eV) of different forms of solid carbon
computed at a given lattice constant $r_0$  (\AA) as a function of the
number of shells  ($N_h$) included in the LR.
The calculations were performed with a TB hamiltonian, with supercells
containing 216, 128 and 100 atoms for diamond, two-dimensional
graphite and the linear chain, respectively.
}
\end{table}
\begin{table}
\begin{tabular}{|c|c|c|c|}
          &         &                &                      \\
Crystal structure   & {\bf $\delta r_0$} ($\%$)   & {\bf $\delta E_c$} ($\%$)
&{\bf $\delta B$} ($\%$)      \\
          &         &                      &                \\ \hline
          &         &                      &                \\
Diamond & 0.2   & 1.4 & 1.0\\
2D-graphite & 0.4   & 2.5 & 1.4\\
1D-chain & 0.5   & 4.7 & 2.7\\
          &         &                &                      \\
\end{tabular}
\vskip 0.5truecm
\caption{
Percentage errors on the
equilibrium lattice parameters  ($\delta r_0$), cohesive energy ($\delta E_c$)
 and bulk modulus  ($\delta B$) of
diamond, graphite and a carbon linear chain, as obtained by
minimizing E[Q] with ${\bf Q}[{\cal N}= 1]$ and $N_h=2$, described within
a TB framework.
The errors  were evaluated with respect to a computation
with extended orbitals.
}
\label{tab2}
\end{table}
\begin{table}
\begin{tabular}{|c|c|c|}
          &                &                    \\
N$_c$
&{\bf $N_h=2$}  & {\bf $N_h= \infty $ }     \\
                   &                &                    \\ \hline
                   &                &                    \\
1-fold & \phantom{4}5     & \phantom{4}4 \\
2-fold & 38    & 42 \\
3-fold & 53    & 50 \\
4-fold & \phantom{4}4     & \phantom{4}4 \\
                   &                &                    \\
\end{tabular}
\vskip 0.5truecm
\caption{
Percentage number of differently coordinated sites ($N_c$)
in liquid C computed as averages over a
TB-MD simulation of 3.8 ps.
The results of the LO formulation with $N_h = 2$ and ${\cal N}=1$
are compared to those of a direct diagonalization scheme
($N_h = \infty$).
}
\label{tab3}
\end{table}
\end{document}